\documentclass[submitting]{nst}
\usepackage{subfigure,dcolumn}
\usepackage{helvet}
\usepackage[english]{babel}
\usepackage[version=4]{mhchem}
\usepackage{amsmath}
\usepackage[x11names]{xcolor}
\usepackage{glossaries}
\usepackage[utf8]{inputenc}
\usepackage{enumitem}
\usepackage{float}
\usepackage{comment}
\usepackage{afterpage}
\usepackage{placeins}
\usepackage{graphicx}
\usepackage{afterpage}
\usepackage{helvet}
\usepackage{courier}
\usepackage{natbib}
\definecolor{DarkGreen}{rgb}{0.0, 0.5, 0.0}

\makenoidxglossaries

\usepackage{listings}
\begin{document}

\title{NON-RESONANT ALPHA-INDUCED NEUTRON-EMISSION: A MULTI- METHOD COMPARISON OF NUCLEAR REACTION RATES}

\author{Bhavay Luthra}
\affiliation{Department of Applied Physics, Delhi Technological University, Delhi, 110042, India}

\author{N. J. Upadhyay}
\affiliation{Department of Physics, Amity School of Applied Sciences, Amity University Maharashtra, Panvel, India}
\email[Corresponding author, ]{bluthra999@gmail.com}

\begin{abstract}
The precise calculation of alpha-induced neutron-emission ($\alpha$,n) reaction rates is fundamental to understanding nucleosynthesis in diverse stellar environments. This study investigates the nuclear reaction rates for various non-resonant alpha-induced neutron-emission reactions by employing analytical, theoretical, and computational methods. We have adopted analytical expressions from existing literature, and for the computational approach, we have implemented the Monte Carlo method. To better understand the functionality and validate the established open-source Monte Carlo scripts, we also developed our own computational code and used it to test various aspects of the method. A comprehensive comparison of the reaction rates from each method is presented by plotting their dependency on temperature.
\end{abstract}


\maketitle

\section{Introduction}

In nuclear astrophysics we are interested in understanding stellar energy production, nucleosynthesis, stellar evolution and stellar explosions, and to do this we need accurate values for reaction cross sections and rates. Examples of this include calculation of the Sun's energy production and estimating the ages of globular clusters.

	Elements till iron are mainly produced from nuclear fusion reactions and elements beyond iron are produced from slow neutron capture (s-process) and rapid neutron capture (r-process) \cite{bib:1}. This paper discusses alpha-induced neutron emission reactions ($\alpha$,n) which occur during late burning stages of massive stars ($M \ge 8 M_o$) and during helium intershell burning in asymptotic giant branch (AGB) stars \cite{bib:2}. 

This is the regime where alpha-induced neutron emission  ($\alpha$,n) reactions channel become a pathway for nucleosynthesis. Reactions such as \ce{^{45}Sc(\alpha,n)^{48}V}, 
     \ \ce{^{51}V(\alpha,n)^{54}Mn}, 
     \ \ce{^{55}Mn(\alpha,n)^{58}Co},
     \ \ce{^{50}Cr(\alpha,n)^{53}Fe} and 
     \ce{^{48}Ti(\alpha,n)^{51}Cr}
 play a key role. They serve two primary functions. First, they act as a source of neutrons within the exploding stellar layers \cite{bib:3}. These neutrons can then fuel further neutron-capture reactions, influencing the final abundances of a wide range of isotopes. Therefore, accurately determining the thermonuclear rates of these ($\alpha$,n) reactions is essential for the predictive power of supernova models and our understanding of the cosmic origin of the elements around us. Second, the intensity and duration of neutron production from ($\alpha$ ,n) reactions determine which path the nucleosynthesis follows at s-process branch points \cite{bib:2, bib:3}.
 
 	The existing well known packages such as TALYS, EMPIRE, and NON-SMOKER are all built Fortran \cite{bib:4,bib:5,bib:6}. These packages computes reaction cross sections and reactions rate by using Hauser-Feshbach statistical model at lower energies which treats average over the many, dense resonances involved in this process and treating reactions non resonant at higher energies \cite{bib:6, bib:7}. In this paper a custom code is developed in python which leverages Monte Carlo sampling by treating reaction as non-resonant. The work also compares it's result by modifying the open source rates-mc which is developed in C++ and also uses monte-carlo sampling \cite{bib:8}. \texttt{rates-mc} was deliberately chosen instead of TALYS, EMPIRE, or NON-SMOKER
 	because rates-mc can isolate resonant reaction rates from non resonant with minor modifications. The goal of the paper is to provide a python based approach to calculate reactions rate for ($\alpha$,n) non resonant reactions.
 \section{Reaction rate formalism}

\textbf{Nuclear Reaction Rates :}
The thermonuclear reaction rate is the number of reactions per unit volume per unit time. For a reaction involving 4 nuclei, 0 and 1 at the entrance channel and nuclei 2 and 3 at the exit channel 
the reaction rate ($r_{01}$) is given by \cite{bib:9}: 

\begin{equation}
r_{01} = N_0 N_1 \langle \sigma v \rangle_{01}
\end{equation}

$N_0$ and $N_1$ are the number densities of the reacting unit and 
$\langle \sigma v \rangle_{01}$ is the reaction rate per particle pair.

The thermally averaged reaction rate per particle pair in units of $cm^3 mol^{-1} s^{-1}$ at the plasma temperature \( T \) in GK is obtained by using relation \cite{bib:10}:

\begin{equation}
\langle \sigma v \rangle_{01} = \left( \frac{8}{\pi m_{01}} \right)^{1/2} 
\frac{1}{(kT)^{3/2}} \int_0^{\infty} E \sigma(E) e^{-E/kT} \, dE
\end{equation}

\( m_{01} \) is the reduced mass in the units of a.m.u and is calculated using $\dfrac{m_0m_1}{m_0+m_1}$, \( k \) is the Boltzmann constant in units of kilo-electron volts per Gigakelvin,  E is the center-of-mass energy in keV, and $\sigma(E)$ is the reaction cross section in barns ($10^{-28} cm^2$). 
 The atomic masses are taken from CIAAW Standard Atomic Weights \cite{bib:11}. Instead of directly substituting energy, maxwell distribution is used to generate velocity. Equation (2) is easily transformed by substituting velocity in place of energy by using $ E = \dfrac{m_{01} v^2}{2}$, yields:
\begin{equation}
\langle \sigma v \rangle_{01} = \left( \frac{8}{\pi m_{01}} \right)^{1/2} 
\frac{1}{(kT)^{3/2}} \int_0^{\infty} \tfrac{m_{01} v^{2}}{2} \sigma(E) e^{-E/kT} \, dE
\end{equation}
The terms in Equation (3) is rearranged to yield the normalized Maxwell-Boltzmann velocity distribution, which is given by:
\begin{equation}
f_T(E)\,dE \;=\; \frac{2}{\sqrt{\pi}} \frac{1}{(kT)^{3/2}} \sqrt{E}\, e^{-E/(kT)} \, dE
\end{equation}

The thermally averaged reaction rate per particle pair in the compact form is written as \cite{bib:12}:
\begin{equation}
	\langle \sigma v \rangle = \int_0^\infty \sigma(E)\, v(E) \, f(E)\, dE
\end{equation}

The \textbf{reaction cross section $\sigma(E)$} is a measure of the  effective area for the reaction that an incoming particle will 
undergo a specified nuclear process when it approaches a target particle.

Determination of the reaction cross-section is experimentally challenging for two main reasons. First, lack of data produced due to very small reactions happening inside the Gamow window and, second due to extreme stellar environment which is difficult to replicate in the laboratory. At lower energies Coulomb repulsion 
between the reacting particles dominates. A nuclei has to tunnel through the barrier in order for reaction to proceed (Quantum tunneling). For most reactions, stellar temperature is too low for nuclei to classically overcome the barrier and quantum tunnelling plays a governing role.
For two particles approaching each other, the actual potential experienced is 
$\propto \dfrac{1}{r} $ (Coulomb potential). To solve it is a computationally heavy task 
due to its complicated shape. The most common approach is to approximate it by dividing it into 
many thin square barriers. The transmission coefficient for the Coulomb potential is then given 
by the product of the transmission coefficients for all of the square barriers. 

For large number of square barriers ($n \to \infty$), the transmission coefficient 
for the Coulomb potential can be found analytically. For very low energies, the transmission coefficient is equal to \cite{bib:9}

\begin{equation}
\hat{T} \approx \exp\!\left(-\frac{2\pi}{\hbar}\,\sqrt{\frac{m_{01}}{2E}}\; Z_0 Z_1 e^2\right)
\end{equation}

This $e^{-2\pi \eta}$ factor is the \textbf{Gamow Factor}, which represents the 
Coulomb barrier tunneling probability. $\eta$ in the Gamow factor is the Sommerfeld parameter and is equal to $\frac{Z_0 Z_1 e^2}{\hbar v} 
= \frac{Z_0 Z_1 e^2}{\hbar} \sqrt{\frac{m_{01}}{2E}}$.  To remove the coulomb barrier penetration probability 
from the reaction cross-section we calculate \textbf{Astrophysical S Factor S(E)}, which is defined by \cite{bib:13}

\begin{equation}
S(E) = E \sigma(E) \exp(2\pi \eta)
\end{equation}

where: $\sigma(E)$ is the reaction cross-section and 
$\exp(2\pi \eta)$ is the inverse of the tunnelling probability factor.\\

We calculate all the different reaction rates by treating reactions as \textbf{non- resonant reactions}. Consequently, the astrophysical factor is weak across the relatively narrow Gamow window. The simplest approximation is to treat the S-factor as constant,
\begin{equation}
S(E) = E \sigma(E) \exp(2\pi \eta) \approx S_0
\end{equation}
To achieve a more accurate description of S-factor, in our modelling S factor is taken to vary smoothly and the general form of S factor is approximated by a polynomial of a degree 4:
\begin{equation}
S(E)  \approx S_4E^4 + S_3E^3 +  S_2E^2 +S_1E +S_0 
\end{equation}
where the $S_i$ are coefficients which are determined from the experimental  and theoretical data. The polynomial is taken from Hussein \emph{et Al.} \cite{bib:14}. The cross section becomes: 

\begin{equation}
\sigma(E) = \dfrac{S(E)}{E \exp(2\pi \eta)}
\end{equation}

Thus, we can use equation (10) and combine it with equation (5)
\begin{equation}
	\langle \sigma v \rangle = \int_a^b \dfrac{S(E)}{E \exp(2\pi \eta)}
\, v(E) \, f(E)\, dE
\end{equation}
For computational efficiency, the limits in the integral in the Equation (11) is replaced by [a,b], from [0,$\infty$].
The reaction probability is maximum in the Gamow window. The limits [a,b] are determined by finding the width ($\Delta$) and the peak ($E_o$) of the Gamow window.
From Equation (11), we define $\phi(E) = \dfrac{E}{kT} + bE^{-1/2} $. To calculate the peak ($E_o$) we differentiate $\phi(E)$ with respect to E and equate it to 0.

\begin{equation}
\frac{d\Phi}{dE} = \frac{1}{kT} - \frac{b}{2} E^{-3/2} = 0 \quad \Rightarrow \quad E_0 = \left( \frac{b k T}{2} \right)^{2/3}
\end{equation}

From Equation (6), \(b \propto Z_0 Z_1\), yielding:
\begin{equation}
E_0 \propto (Z_1 Z_2)^{2/3} T^{2/3}
\end{equation}
The width ($\Delta$) is calculated, by performing Taylor expansion of \(\Phi(E)\) around $E_o$ to second order:

\begin{align}
\Phi(E) &\approx \Phi(E_0) + \frac{1}{2} \Phi''(E_0) (E - E_0)^2 \\
\Phi''(E) &= \frac{3b}{4} E_0^{-5/2}
\end{align}
 The value of b is used from Equation (11). $\Phi''(E_0)$ is given by

\begin{equation}
	\Phi''(E_0) = \frac{3}{2} \cdot \frac{1}{E_0} \cdot \frac{1}{kT} = \frac{3}{2 E_0 kT}
\end{equation}

This allows the integrand's exponential term to be approximated as a Gaussian function centered at 
$E_o$
\begin{equation}
e^{-\Phi(E)} \approx e^{-\Phi(E_0)} \exp\left[-\frac{1}{2} \Phi''(E_0) (E - E_0)^2\right]
\end{equation}
The standard Gaussian function is given by:
\begin{equation}
	\exp\left[-\frac{(E - E_0)^2}{(\Delta/2)^2}\right]
\end{equation}

We compare the curvature's peak in Equation (18) with our integrand, giving:

\begin{equation}
	\frac{1}{(\Delta/2)^2} = \frac{1}{2} \Phi''(E_0) \quad \Rightarrow \quad \Delta = \frac{4}{\sqrt{3}} \sqrt{E_0 kT}
\end{equation}

\section{Methodology}
This subsection discusses implementation of the monte-carlo method, structure of input data and explains how \texttt{rates-mc} is modified to solely calculate reaction rate for non-resonant reaction.
\begin{enumerate}
\item \textbf{Our script:} A Python script is developed to calculate thermally averaged reaction rate per particle pair as described in Equation (11). We use NumPy and SciPy to perform numerical calculation and Matplotlib and pandas for data handling and visualization. In this work, we only present the median reaction rate for $(\alpha,n)$ reactions.  We implement the code using the Monte-Carlo method for extensibility. A common class \texttt{Reaction} is defined 
for all reactions. The input parameters and its instance methods are described in Table 1 and Table 2 respectively.

The \texttt{\_integrand} instance method encapsulate the steps to compute integrand described in Equation (5), such as evaluating the S-factor (Equation (9)), tunnelling factor (Equation (6)), and the velocity of the nuclei for a particular value of energy. After setting up the integrand we calculate the integral using the \texttt{scipy.Quad} method and the Monte-Carlo method. Both methods use the same integral limits. The lower integration limit is the coulombic repulsion  (\texttt{E\_th\_KeV}). To decrease the computation time, instead of defining another instance method which dynamically  calculates the Gamow peak and width for each reaction, a common upper integration limit is defined as a function of temperate which is given by:
\begin{equation}
	E\_th\_keV + 25 kT\_keV
\end{equation}
 This is justified as Beyond this point the Maxwell–Boltzmann distribution is suppressed by 
$e^{(-25)} \approx 1.4 \times 10^{-11}$, so contributions from higher energies are negligible at numerical precision.
At last, we calculate the thermally averaged reaction rate using both methods. 
\item \textbf{Modification of \texttt{rates-mc}:} By default, the \texttt{rates-mc} computes reactions for resonant reactions even if the input parameters are given for non-resonant reactions.  We completely disabled the resonant calculation part by modifying the \texttt{CMakeCache.txt}.
\end{enumerate}

Finally, result is then plotted and compared using the all 3 methods.

\begin{table}[!htb]
\caption{Parameters of \texttt{Reaction} class}
\label{tab:parameters}
\centering
\begin{tabular}{ll}
\toprule
\textbf{Parameters} & \textbf{Description} \\
\midrule
  label    &  Label for the ($\alpha$,n) reactions \\
Z1       & Charge of the target nuclei \\
mass\_target\_amu       & Charge of the target nuclei \\
 E\_th\_keV & Minimum energy of nuclei particles\\
coeffs  & Coefficients for S-factor \\
\bottomrule
\end{tabular}
\end{table}

\begin{table}[!htb]
\centering
\caption{Instance methods of \protect\texttt{Reaction} class}
\label{tab:methods}
\begin{tabular}{lp{6cm}}
\toprule
\textbf{Parameters} & \textbf{Description} \\
\midrule
\texttt{lnS} & Calculate the S-factor \\
\texttt{\_integrand} & Sets up the integrand \\
\texttt{calculate\_rates} & Calculate the integration using adaptive Gauss--Kronrod quadrature method \\
\texttt{calculate\_rates\_mc} & Calculate the integration using the Monte--Carlo method \\
\bottomrule
\end{tabular}
\end{table}

\section{Results}

In this section, figures are presented showing the temperature dependency of reaction rate per particle pair on temperature for all 5 $(\alpha$,n) reactions from the previously described three methods in methodology section.

The results obtained from the the modified \texttt{rates-mc} code, and the custom-developed Python Monte Carlo and \texttt{scipy.Quad} method code are compared for all five ($\alpha$,n) reactions. For each reaction, the reaction rate per particle pair $\langle \sigma v \rangle$ is plotted as a function of temperature between 0.1 GK and 5 GK. The comparison is shown in Figures 1–5.

For each case, all three methods exhibit similar temperature-dependent trends, with  $\langle \sigma v \rangle$  increasing sharply with temperature as expected from the enhanced tunnelling probability through the Coulomb barrier. The Monte Carlo and \texttt{scipy.Quad} completely overlap each other in all five reactions across all of the temperature range, indicating that the numerical integration is correctly performed using Monte Carlo method.

 Small deviations ($<$ 5\%) are observed at low temperatures (T $<$ 2 GK), where the rate changes most rapidly, likely due to the limited number of Monte Carlo samples and numerical integration sensitivity.

\begin{figure*}[htbp!]
  \centering
  \includegraphics[width=0.9\textwidth]{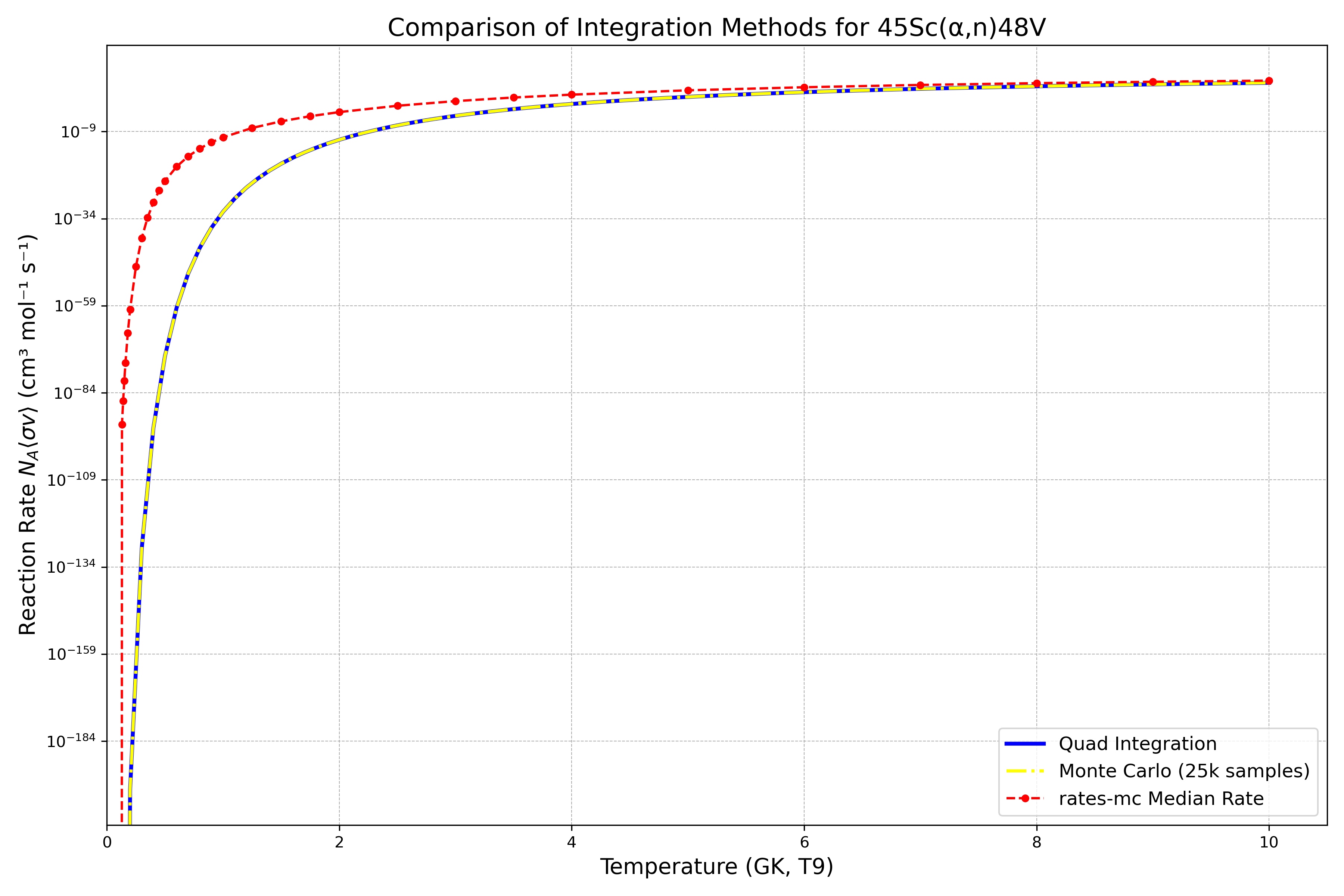}
  \caption{Comparison of experimental and calculated cross sections for the 
  \(^{45}\)Sc(\(\alpha,n\))\(^{48}\)V reaction.}
  \label{fig:sc_v}
\end{figure*}

\begin{figure*}[htbp!]
  \centering
  \includegraphics[width=0.9\textwidth]{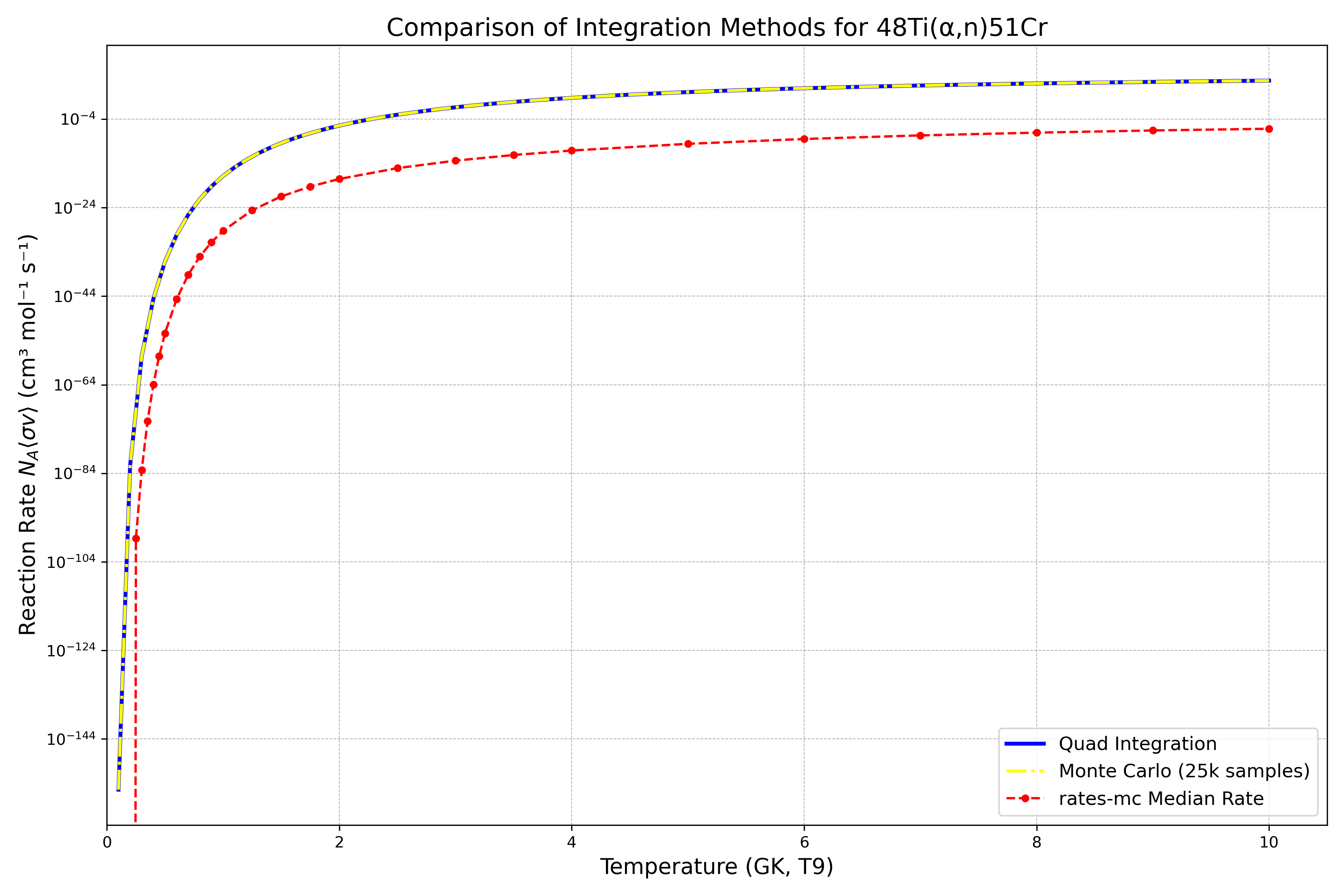}
  \caption{Comparison of experimental and calculated cross sections for the 
  \(^{48}\)Ti(\(\alpha,n\))\(^{51}\)Cr reaction.}
  \label{fig:ti_cr}
\end{figure*}

\begin{figure*}[htbp!]
  \centering
  \includegraphics[width=0.9\textwidth]{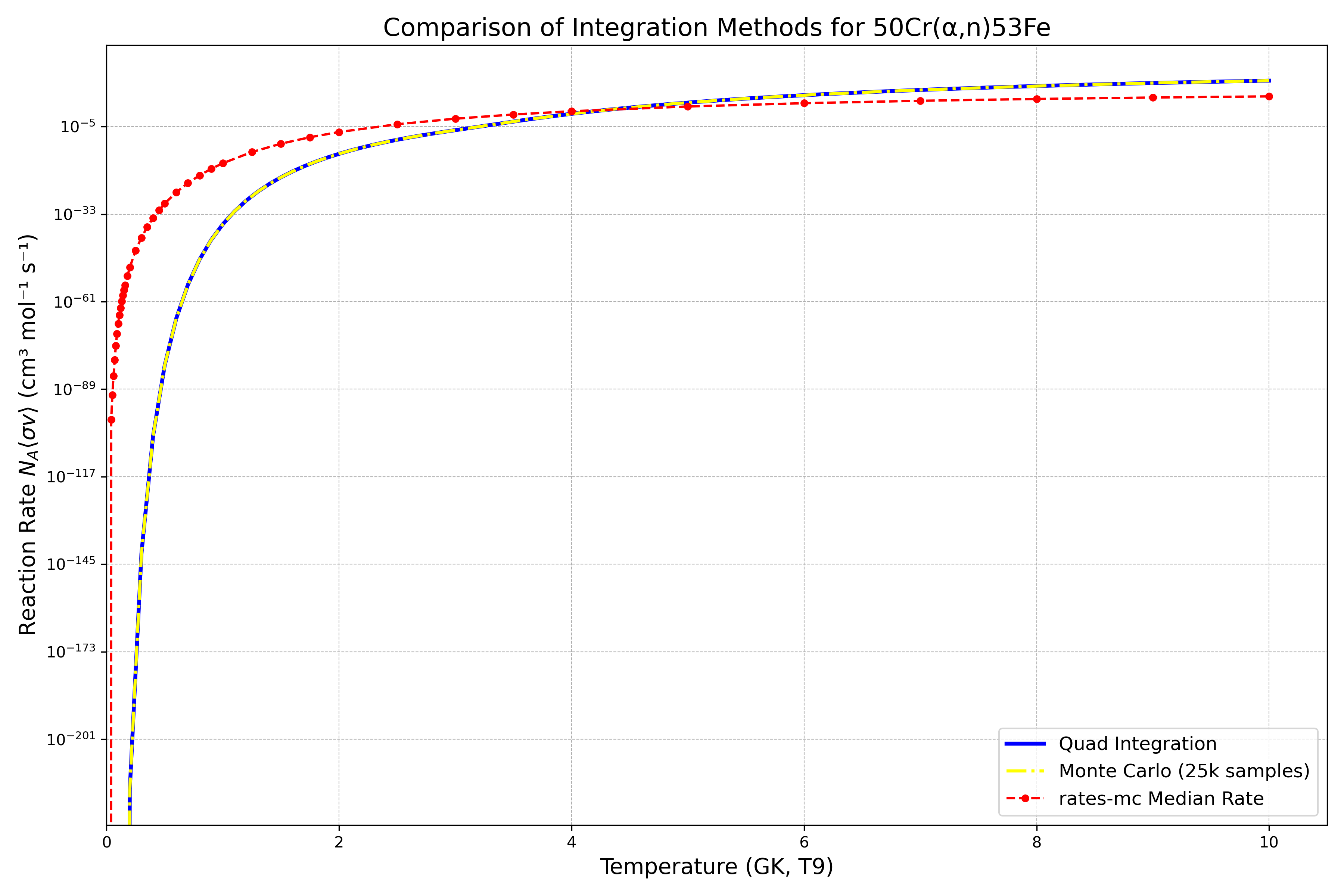}
  \caption{Comparison of experimental and calculated cross sections for the 
  \(^{50}\)Cr(\(\alpha,n\))\(^{53}\)Fe reaction.}
  \label{fig:cr_fe}
\end{figure*}

\begin{figure*}[htbp!]
  \centering
  \includegraphics[width=0.9\textwidth]{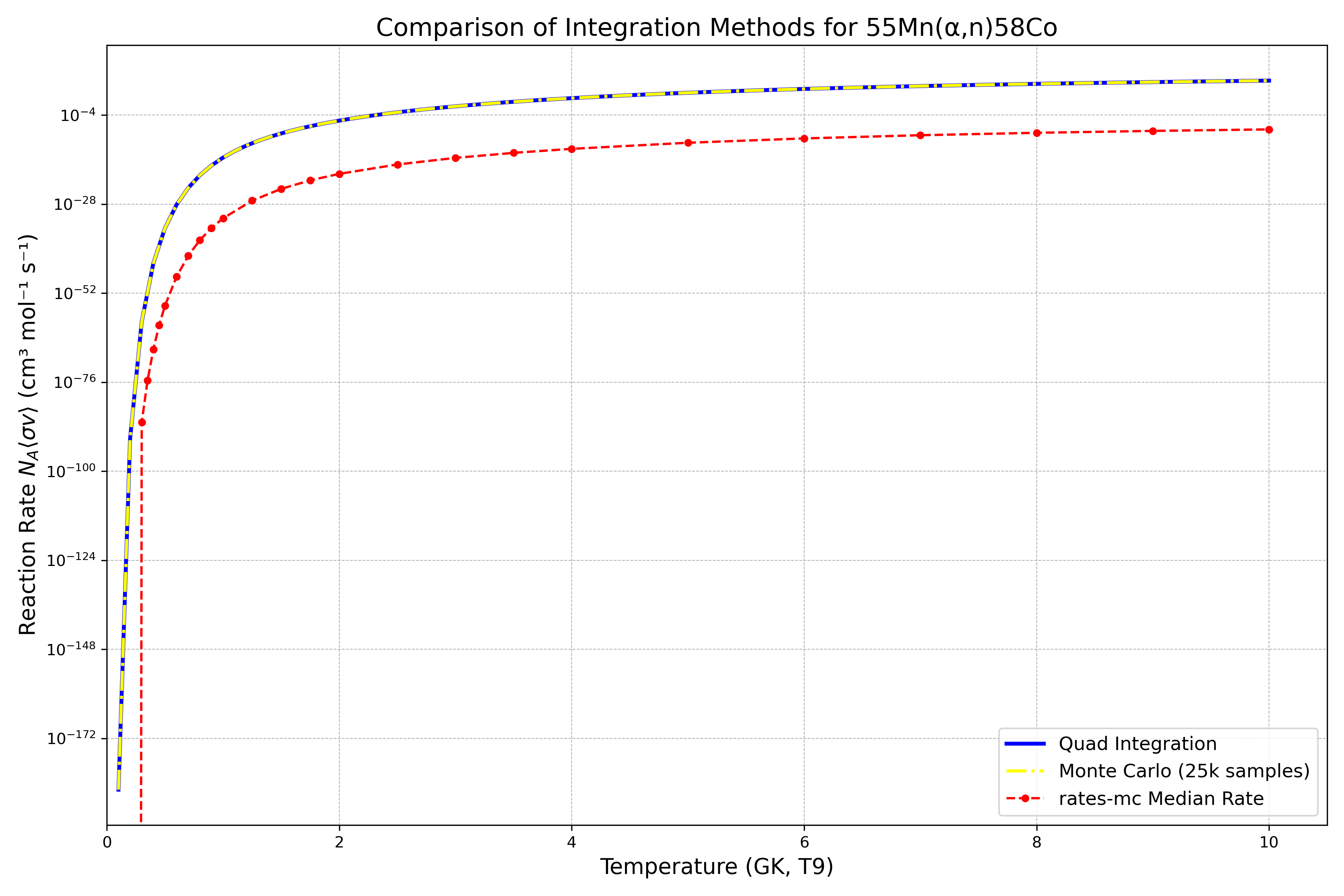}
  \caption{Comparison of experimental and calculated cross sections for the 
  \(^{55}\)Mn(\(\alpha,n\))\(^{58}\)Co reaction.}
  \label{fig:mn_co}
\end{figure*}

\begin{figure*}[htbp!]
  \centering
  \includegraphics[width=0.9\textwidth]{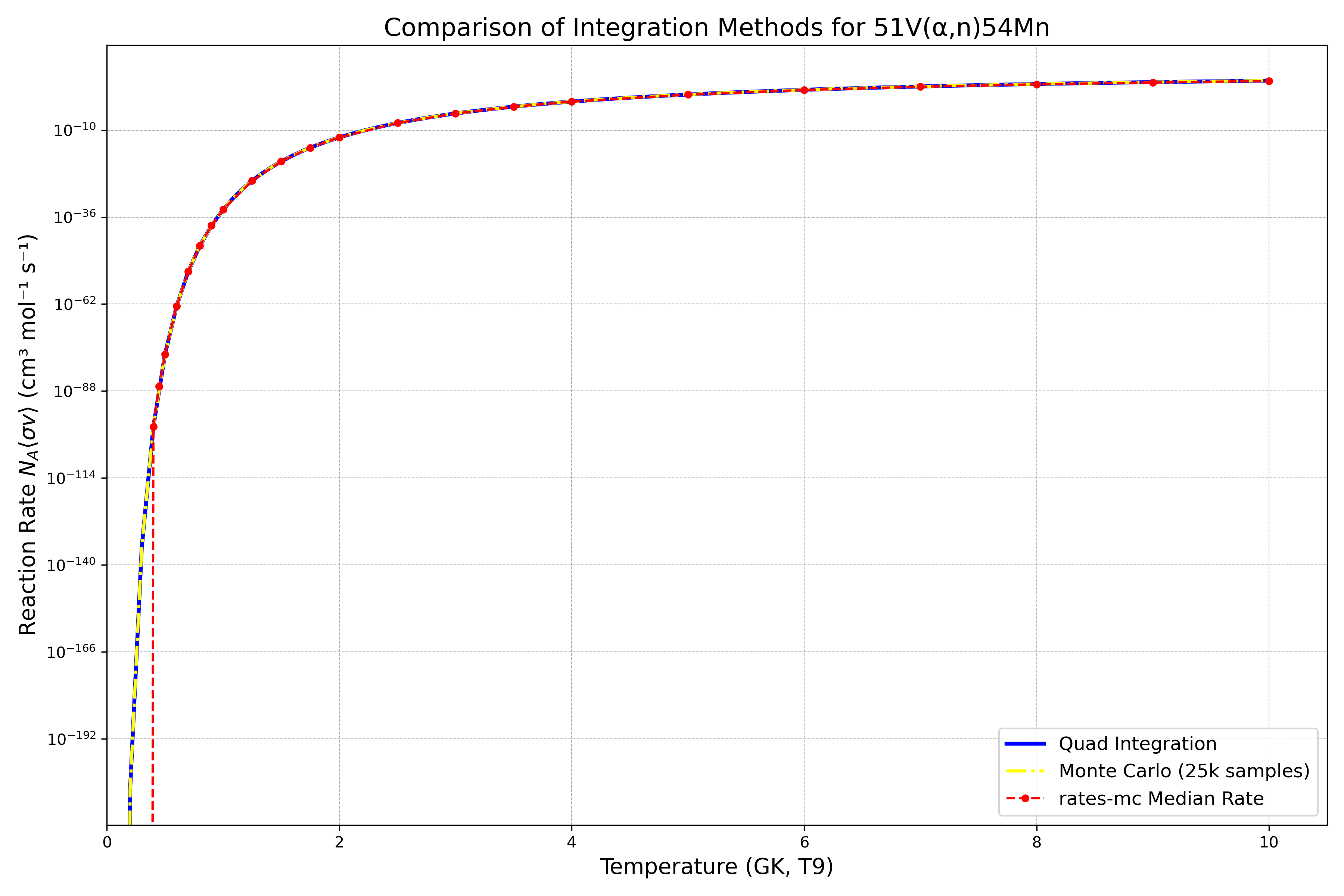}
  \caption{Comparison of experimental and calculated cross sections for the 
  \(^{51}\)V(\(\alpha,n\))\(^{54}\)Mn reaction.}
  \label{fig:v_mn}
\end{figure*}

\section{Conclusion}
In this paper, the non-resonant reaction rates of five different alpha-induced neutron-emission ($\alpha$,n) reactions are calculated and presented. The reaction rates are calculated by three different methods.
The results from all three methods confirm a strong temperature dependency for each reaction, which is presented by plotting their dependency on temperature. The shape of graphs is identical for all the five reaction by all three methods. , Slight deviations between the methods are primarily observed at low temperatures (T < 2 GK) where the rate is changing most rapidly.
 This work currently presents only the median reaction rate. Calculating the reaction uncertainty is beyond its scope and will be pursued in future work.

\newpage

\bibliography{references}

\end{document}